\documentclass[submission,copyright,creativecommons]{eptcs}
\providecommand{\event}{CompSust 2012} 
\usepackage{amssymb,amsmath,url}
\usepackage{breakurl}
\usepackage{custom}
\usepackage{algorithmic}
\usepackage{algorithm}
\usepackage{graphicx}
\usepackage{epstopdf}
\usepackage{epsf}
\usepackage{comment} 
\usepackage{titlesec}

\graphicspath{{./images/}{E:/GRADUA~1/Wortman/Thesis/BTree/CSUF-T~1/images/}}
\titleformat{\section}{\large\bfseries}{\thesection}{1em}{\vspace{-1mm}}
 
\title{A Durable Flash Memory Search Tree}
\author{
James Clay III \qquad\qquad Kevin Wortman
\institute{California State Univeristy, Fullerton}
\email{\quad jaclay@fullerton.edu \quad\qquad kwortman@fullerton.edu}
}
\def\titlerunning{An Efficient Flash Based B-Tree}
\def\authorrunning{J. Clay III, K. Wortman}
\begin{document}
\maketitle

\begin{abstract}
We consider the task of optimizing the \emph{B-tree} data structure, used extensively in operating systems and databases, 
for sustainable usage on multi-level flash memory. 
Empirical evidence shows that this new flash memory tree, or \emph{FM Tree,} extends the operational lifespan of each block of flash memory by a factor of roughly 27 to 70 times, while still supporting logarithmic-time search tree operations.
\end{abstract}

\section{Introduction}

Flash memory has seen growing usage in recent years across all areas of computing technology;
devices utilizing flash include tablets, cell phones, and USB drives. The lifespan of flash memories has
been of serious concern since their inception; flash memory degrades proportionally to the number of
times it is erased. While significant advances have been made to combat this effect, flash usage continues
to grow in areas where rewrite-intensive operations are necessary. We consider the problem of extending the operational lifetime of flash memory by avoiding erase operations, at the expense of some additional time and space overhead.

\section{Background}
\subsection{Flash Memory}
\emph{Multi-level flash memory} is a form of NAND random access memory \emph{(RAM)} capable of storing one of $q>1$ discrete states in each flash cell. These $q$ values may interpreted as nonnegative integers in the range $[0, \, q-1].$
Cells are aggregated into \emph{blocks} of fixed size. Individual multi-level flash cells may be incremented, which is a fast and non-destructive operation. However, idiosyncratically, bucket states can only be decreased by resetting an entire block to state 0 \emph{en masse.} These \emph{resets} or \emph{erasures} are costly both in terms of time of the operation
and lifetime of the device. Typical block erases take between 1.5-2 milliseconds in
comparison to seek or write times which are in the tens or hundreds of microseconds.
Each NAND flash device has a set number of expected erasure cycles it can
perform before failing. Wear distribution can be done in multiple
ways ranging from a purely round robin approach to keeping the most actively used
blocks in RAM. However recent advances in multiple level flash memory data
representation relaxes the requirement that a block must be erased before it can be
rewritten. Since erasures are a limiting factor to both the durability and write throughput of multi-level flash memory, we investigate approaches to avoid erasures, at the cost of modest constant-factor expenditures of space and CPU time.
 
 \subsection{Tree Types}
Our FM tree is an amalgam of ideas from established search tree data structures. In this section we survey their properties.
A \emph{Bayer McCreight  B-Tree,} henceforth denoted \emph{B-tree}, 
is a type of balanced search tree developed for managing large blocks of
data, particularly in file-systems and databases.
A B$^+$ tree resembles a B-tree, with the addition of redundant node links that facilitate tree traversal in common database operations.
B$^-$ trees are a relaxed version of the common B$^+$ tree where the notoriously complex post-deletion rebalancing operation is omitted. Perhaps surprisingly, Sen and Tarjan showed that, despite postponing rebalance operations, B$^-$ trees still boast asymptotically optimal update operations up to amortization, and may be implemented simply with attractive constant factors.

\subsection{A Durable B-tree}

In creating a durable B-tree we have made a variety of changes to 
the implementation of both operations on and storage of the keys within the B-tree. These alterations reduce the maximum number of erasures per block and the average number of erasures across all blocks.
\begin{itemize}
\item Block erasures are performed lazily, postponing them for as long as possible.
\item The requirement that key/value pairs be sorted within nodes is relaxed. This gives the insertion operation leeway to reuse key/value slots without erasing the block, but makes searching a single node take $O(B)$ rather than $O(\log B)$ time.
\item As in the B$^-$ tree, the delete operation marks unused nodes \emph{barren} rather than actually splicing the node out of the tree. Barren nodes are ignored until the tree is eventually garbage collected and rebuilt. This allows a node to be excised by toggling flash cell(s) representing a boolean barren flag, without performing block erasures.
\end{itemize}

\section{Analysis}

We prove that an FM tree supports the search, insert, and delete operations all in amortized $O(\log n)$ time, matching the lower bound for amortized search tree data structures. We show that, for any sequence of operations, an FM tree performs strictly fewer erasures than a conventional B-tree.
 
\section{Experimental Results}

We present a variety of experiments performed on a Python implementation of the FM Tree.
We emulate the flash memory, FM Tree, and B-tree to run a variety of benchmarks.
Every experimental trial consists of randomly generated data sets that are inserted into both trees. 
Each tree is inserted with a baseline of 1000 elements. 
Following these initial insertions, 10000 randomly chosen insertions and deletions are performed.
We repeat this process a total of 4 times independently and determine the average for each data point. 
We then calculate the FM Tree performance by comparing the erasures, reads and writes between it and the B-tree.
This process indicates that the FM Tree performs 27 times to 72.2 times fewer erasures.
While the total read count was higher, the total writes and erasures performed were far lower. 
As these are the most expensive operations in terms of time, realistically the FM Tree would also be far faster than the B-tree.

\section{Conclusion}

We find that the FM-Tree is a more durable, faster variant of the B-tree with 
properties that make it intrinsically better for operating on flash memory. 
We also show that the erasure count for the FM Tree is drastically smaller (27 to 72 times) than that of a B-tree.
 
\end{document}